\documentclass[aps,prc,letterpaper,11pt,twoside,tightenlines,nofootinbib,showpacs,
preprint,superscriptaddress]{revtex4-1}
\usepackage{graphicx}
\usepackage[sort&compress]{natbib}
\usepackage{subfigure}
\usepackage{amsmath}
\usepackage{amsfonts}
\usepackage{cancel}

\newcommand{\mev}{\textrm{ MeV}}
\newcommand{\be}{\begin{equation}}
\newcommand{\ee}{\end{equation}}
\newcommand{\ba}{\begin{eqnarray}}
\newcommand{\ea}{\end{eqnarray}}
\newcommand{\gev}{\textrm{ GeV}}

\begin{document}

\title{A study of $\eta K \bar K$ and $\eta' K \bar K$ with the fixed center approximation to Faddeev equations.}

%\author{Weihong Liang$\,^{1,\;2,}$, C. W. Xiao$\,^2$, and E. Oset$\,^2$}

%\affiliation{
%$^1$ Department of Physics, Guangxi Normal University, Guilin, 541004, P. R. China \\
%$^2$ Departamento de F\'{\i}sica Te\'orica and IFIC, Centro Mixto Universidad \\de Valencia-CSIC, %Institutos de Investigaci\'on de Paterna, Apartado 22085, 46071 Valencia, Spain}

\author{Weihong Liang}
\email{liangwh@gxnu.edu.cn}
\affiliation{
Department of Physics, Guangxi Normal University, Guilin, 541004, P. R. China}
\affiliation{
Departamento de F\'{\i}sica Te\'orica and IFIC, Centro Mixto Universidad \\de Valencia-CSIC, Institutos de Investigaci\'on de Paterna, Apartado 22085, 46071 Valencia, Spain}

\author{C. W. Xiao}
\author{E. Oset}
\affiliation{
Departamento de F\'{\i}sica Te\'orica and IFIC, Centro Mixto Universidad \\de Valencia-CSIC, Institutos de Investigaci\'on de Paterna, Apartado 22085, 46071 Valencia, Spain}

\date{\today}

\begin{abstract}

In the present work we investigate the three-body systems of $\eta K \bar K$ and $\eta' K \bar K$, by taking the fixed center approximation to Faddeev equations. We find a clear and stable resonance structure around 1490 MeV in the squared $\eta K \bar K$ scattering amplitude, which is not sensitive to the renormalization parameters. Conversely, we get only an enhancement effect of the threshold in the $\eta' K \bar K$ amplitude that indicates the difficulty to bind the  $\eta' K \bar K$ system as a consequence of a weaker  $\eta' K$ interaction than the $\eta K$ one.   We  associate the $\eta K \bar K$ state found to the $\eta(1475)$.
\end{abstract}


\maketitle

\section{Introduction}

Understanding the nature and structure of hadronic resonances is a main topic in high energy physics, which attracts the attention of both theory and experiment. With the advent of quantum chromodynamics (QCD) and the standard model, modern hadron physics is developing fast. The traditional picture for the internal structure of hadrons is that a meson is made of $q \bar q$ and a baryon of $qqq$, and quark models describe them well. On the other hand, with the development of the experiments, some states have been found experimentally whose properties cannot be explained by the standard way and may be of more complex structures, like tetraquarks and hybrids including possible glueballs for mesons, pentaquarks and heptaquarks for the baryons, or molecular states (see recent reviews in Refs. \cite{Crede:2013kia,Klempt:2009pi}). For the low energy region where the abnormal states showed up, nonperturbative QCD should be explored, such as Lattice QCD \cite{Kogut:1982ds,Luscher:1996sc,Luscher:1996ug,Mohler:2012na}, QCD sum rule \cite{Shifman:1978bx,Reinders:1984sr,Dias:2012ek,Hidalgo-Duque:2013pva}, effective field theory \cite{Politzer:1988bs,Georgi:1990um,Epelbaum:2008ga}, Dyson-Schwinger equations \cite{Roberts:1994dr,Maris:2003vk,Fischer:2006ub}, chiral quark model \cite{Manohar:1983md,Zhang:1997ny}, chiral unitary approach \cite{Kaiser:1995cy,Oller:2000ma,Oller:2000fj,Hyodo:2008xr,Oset:2008qh}, and so on. Chiral dynamics for meson-meson and meson-baryon interaction has played an important role in understanding the nature and structure of hadronic resonances, and it has shown that many known resonances are generated dynamically as a natural consequence of the hadron-hadron interaction, much as the deuteron appears as a simple bound state of a proton and a neutron.

Following the spirit of the approach of Refs. \cite{Kaiser:1995cy,Kaiser:1995eg}, in Ref. \cite{Oller:1997ti} the kernel of the interaction (potential) of pseudoscalar mesons was evaluated starting from the chiral Lagrangians \cite{Gasser:1984gg, Meissner:1993ah,Pich:1995bw,Ecker:1994gg,Bernard:1995dp}, and then, implementing unitarity in coupled channels, the scalar meson resonances $\sigma$ [or $f_0(500)$], $f_0(980)$, $a_0(980)$ were dynamically produced, with phase shifts, inelasticities, mass distributions of given channels consistent with the experimental data (the consistency of the results with QCD sum rules can be seen in Refs. \cite{Chen:2007xr,Liu:2006xja}). Along the same lines, the unitarity approach with the coupled channels explains successfully both the experimental data for the light scalar mesons \cite{Oller:1997ti,Oller:1998hw,Dai:2011bs} (such as $a_0(980),~f_0(980),~\sigma$,~ and ~$\kappa$~ [or $K^*(800)$]) and the light baryons \cite{Kaiser:1995eg,Oset:1997it,Oller:2000fj,Inoue:2001ip,Jido:2003cb,GarciaRecio:2002td,Hyodo:2002pk}, two $\Lambda(1405),~\Lambda(1670),~N^*(1535),~\Delta(1620)$, etc. Extrapolation of this dynamics to the charm sector has also produced many meson states, as the $D^*_{s0}(2317),~D^*_0(2400),~X(3700),~X(3872)$, etc \cite{kolo,hofmann,Guo:2006fu,Gamermann:2006nm,danielax}, as well as baryon states like the $\Lambda_c(2595)$ \cite{hofmann2,Mizutani:2006vq,Tolos:2007vh}. More work on the $K \pi$ interaction is done in Refs. \cite{Oller:1998zr,Li:2002we,Guo:2005wp}, where starting from the chiral Lagrangian and taking into account unitarity, the $S$-wave $K \pi$ elastic scattering amplitude is evaluated and good agreement with the experimental phase shifts is obtained. In addition, the scalar resonance $\kappa$ is generated dynamically, which is also seen in the final state interaction in some reactions \cite{Oller:2004xm}.

The three-body interaction is another subject in the hadron physics which is also drawing much attention for a long time \cite{Fujita:1957zz,Faddeev:1960su,Glockle:1986zz,Weinberg:1992yk,Richard:1992uk}.  Combining the three-body Faddeev equations with chiral dynamics, Ref. \cite{alberone} has reported several $S$-wave $J^P=\frac{1}{2}^+$ resonances which qualify as two mesons-one baryon molecular states. This combination of Faddeev equations and chiral dynamics produces results consistent with QCD sum rules in the investigation of the $D K \bar K$ system in the work of Ref. \cite{MartinezTorres:2012jr}. On the other hand, by taking the fixed center approximation (FCA) \cite{Faddeev:1960su,Toker:1981zh,Barrett:1999cw,Deloff:1999gc,Kamalov:2000iy,Gal:2006cw} to Faddeev equations, several multi-$\rho(770)$ states are dynamically produced in Ref. \cite{Roca2010}, in which the resonances $f_2(1270)(2^{++})$, $\rho_3(1690)(3^{--})$, $f_4(2050)(4^{++})$, $\rho_5(2350)(5^{--})$, and $f_6(2510)(6^{++})$ are theoretically found as basically molecules of an increasing number of $\rho(770)$ particles with parallel spins. Analogously, in Ref. \cite{Yamagata2010}, the resonances $K^*_2(1430)$, $K^*_3(1780)$, $K^*_4(2045)$, $K^*_5(2380)$ and a new $K^*_6$ could be explained as molecules with the components of an increasing number of $\rho(770)$ and one $K^*(892)$ meson. In the same direction, the work of Ref. \cite{Xiao:2012dw} predicts several charmed resonances, $D^*_3$, $D^*_4$, $D^*_5$ and $D^*_6$. With the same approach, the $\Delta_{\frac{5}{2}^+}(2000)$ puzzle is solved in Ref. \cite{Xie:2011uw} in the study of the $\pi-(\Delta\rho)$ interaction. The FCA to Faddeev equations is technically simple, and allows one to deal with three-body hadron interactions which would be otherwise rather cumbersome \cite{Xie:2010ig,Bayar:2011qj,Xiao:2011rc,Bayar:2012rk}. As discussed in Ref. \cite{Bayar:2011qj}, this method is accurate when dealing with bound states,  and gets consistent results with the full Faddeev equation evaluation without taking FCA \cite{MartinezTorres:2008kh}, or a variational calculation with  a nonrelativistic three-body potential model \cite{Jido:2008kp} (more discussions can be seen in Ref. \cite{Oset:2012gi}). One should also know the limits of the applicability of the FCA, and one should avoid the case in which the states have enough energy to excite its components in intermediate states \cite{MartinezTorres:2010ax}, which is the case of resonances above threshold. Recently, the results of the FCA to Faddeev equations have been confirmed by the variational method approach with the effective one-channel Hamiltonian in Ref. \cite{Bayar:2012dd}, which uses the two methods to study the $DNN$ system and predicts a narrow quasi-bound state with the mass of about 3500 MeV.

In our present work we will use the FCA to Faddeev equations to investigate the  $\eta K \bar K$ and $\eta' K \bar K$ systems. When studied in $S$-wave, provided the strength of the interaction allows for it, this systems could give rise to $\eta$ states. There are many $\eta$ excited states, the lowest ones the $\eta(1295)$,  $\eta(1405)$ and $\eta(1475)$. Since we do not want states too far from threshold, the $\eta(1475)$ could be in principle a candidate for the $\eta K \bar K$ system. For the $\eta ' K \bar K$ system we would have to look for an $\eta$ state around 1930 MeV. There are two $\eta$ states around this energy in the Particle Data Group (PDG) \cite{pdg2012}, the $\eta(1760)$ and the $\eta(2225)$, both far away from the $\eta ' K \bar K$ threshold mass. There is a peak seen at 1870 MeV in the $J/\psi \to \eta \pi^+ \pi^-$ in Ref. \cite{Ablikim:2011pu}, but its quantum numbers are not well determined. Similarly, there is another peak seen in the
$J/\psi \to \eta '\pi^+ \pi^-$ reaction that peaks around 1836 MeV [$X(1835)$], with a large width of about 190 MeV \cite{Ablikim:2010au}. We shall explore the possibility that the $\eta ' K \bar K$ could be responsible for any of such states, although we anticipate that the interaction is too weak to lead to such strongly bound systems.

The $\eta  K \bar K$ and $\eta ' K \bar K$ systems have been investigated before in Ref. \cite{migueleta}, following the lines of Ref. \cite{AlvarezRuso:2009xn}, where it was concluded that the $\eta  K \bar K$ system could be the $\eta(1475)$ resonance, and the $\eta ' K \bar K$ the $X(1835)$.  Yet, in Ref. \cite{MartinezTorres:2010ax} it was discussed that the method of Ref. \cite{AlvarezRuso:2009xn} contains some element of uncertainty which makes it most opportune to perform calculations with a different method and contrast the predictions. Certainly, there are also other options for these resonances using quark models and other approaches and a detailed discussion on it can be found in the Introduction of Ref. \cite{migueleta}. In the present paper we will explore the possible molecular structure of these three body systems.

\section{Multi-body interaction formalism}

\begin{figure}
\centering
\includegraphics[scale=0.4]{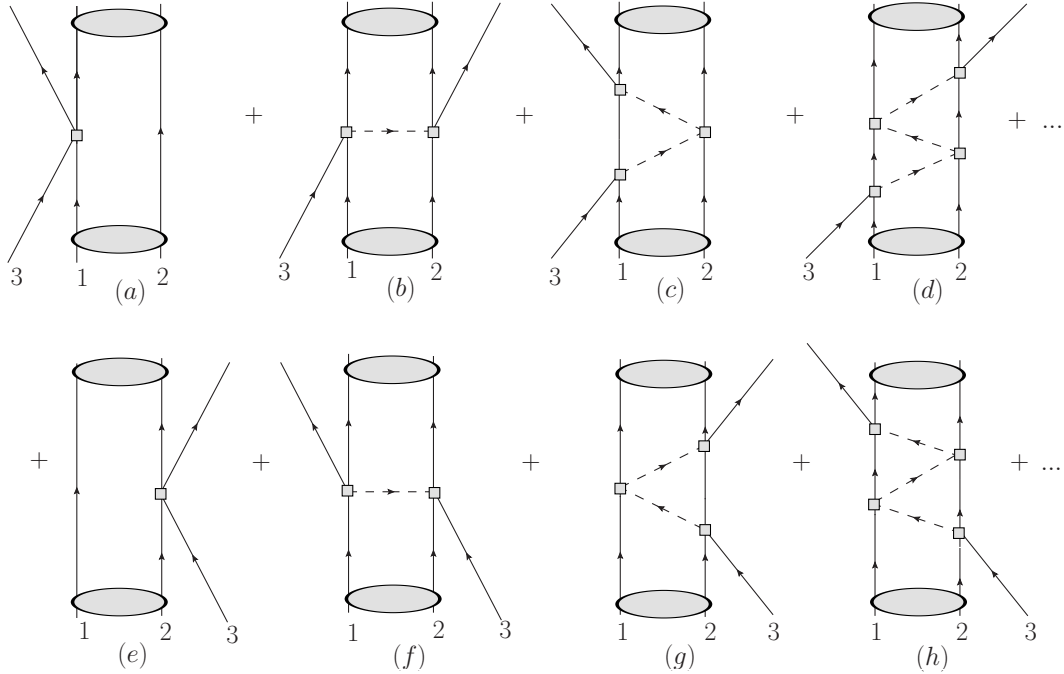}
\caption{Diagrammatic representation of the FCA to Faddeev equations.}\label{FCAfig}
\end{figure}

The Faddeev equations under the FCA are an effective tool to deal with multi-hadron interaction \cite{Toker:1981zh,Barrett:1999cw,Deloff:1999gc,Kamalov:2000iy,Gal:2006cw,Roca2010,Yamagata2010,Xiao:2012dw,Xie:2011uw,Xie:2010ig,Bayar:2011qj,Xiao:2011rc, Bayar:2012rk}. They are particularly suited to study systems in which a pair of particles cluster together and the cluster is not much modified by the third particle. The FCA to Faddeev equations assumes a pair of particles (1 and 2) forming a cluster. Then particle 3 interacts with the components of the cluster, undergoing all possible multiple scattering with those components. This is depicted in Fig. \ref{FCAfig}.
With this basic idea of the FCA, we can write the Faddeev equations easily. For this one defines two partition functions $T_1$ and $T_2$, which sum all diagrams of the series of Fig. \ref{FCAfig} which begin with the interaction of particle 3 with particle 1 of the cluster ($T_1$), or with the particle 2 ($T_2$). The equations then read
\begin{align}
T_1&=t_1+t_1G_0T_2,\\
T_2&=t_2+t_2G_0T_1,\\
T&=T_1+T_2,
\end{align}
where $T$ is the total three-body scattering amplitude that we are looking for. The amplitudes $t_1$ and $t_2$ represent the unitary scattering amplitudes with coupled channels for the interactions of particle 3 with particle 1 and 2, respectively. And $G_0$ is the propagator of particle 3 between the components of the two-body system. In our case we will take the $K \bar K$ forming a cluster of the $f_0(980)$ (we shall discuss the $a_0(980)$ also), as appears in the chiral unitary approach \cite{Oller:1997ti,kaiser,markushin,juanito,rios} and the $\eta$ or $\eta '$ will be the particle 3.

For the unitary amplitudes corresponding to single-scattering contribution, one must take into account the isospin structure of the cluster and write the $t_1$ and $t_2$ amplitudes in terms of the isospin amplitudes of the (3,1) and (3,2) systems. But this is trivial in the present case since the $\eta$, $\eta '$ have I=0 and hence the $\eta K$, $\eta ' K$ are in I=1/2. Besides, because of the normalization of Mandl and Shaw \cite{mandl} which has different weight factors for the particle fields, we must take into account how these factors appear in the single scattering and double scattering and in the total amplitude. This is easy and is done in detail in Refs. \cite{Yamagata2010,Xie:2010ig}. We show below the details for the present case of a meson cluster (also particle 1 and 2) and a meson as scattering particle (the third particle). In this case, following the field normalization of Ref. \cite{mandl} we find for the $S$ matrix of single scattering,
\begin{align}
S^{(1)}_1=&-it_1 (2\pi)^4\,\delta(k+k_R-k'-k'_R) \times \frac{1}{{\cal V}^2}
 \frac{1}{\sqrt{2\omega_3}} \frac{1}{\sqrt{2\omega'_3}}
 \frac{1}{\sqrt{2\omega_1}} \frac{1}{\sqrt{2\omega'_1}},\label{s11}\\
S^{(1)}_2=&-it_2 (2\pi)^4\,\delta(k+k_R-k'-k'_R) \times\frac{1}{{\cal V}^2}
 \frac{1}{\sqrt{2\omega_3}} \frac{1}{\sqrt{2\omega'_3}}
 \frac{1}{\sqrt{2\omega_2}} \frac{1}{\sqrt{2\omega'_2}},\label{s12}
\end{align}
where, $k,\,k'$ ($k_R,\,k'_R$) refer to the momentum of initial, final scattering particle ($R$ for the cluster), $\omega_i,\,\omega'_i$ are the energies of the initial, final particles,  $\cal V$ is the volume of the box where the states are normalized to unity and the subscripts 1, 2 refer to scattering with particle 1 or 2 of the cluster.

  The double scattering diagram, Fig. \ref{FCAfig} (b), is given by
\be
\begin{split}
S^{(2)}=&-i(2\pi)^4 \delta(k+k_R-k'-k'_R) \frac{1}{{\cal V}^2}
\frac{1}{\sqrt{2\omega_3}} \frac{1}{\sqrt{2\omega'_3}}
 \frac{1}{\sqrt{2\omega_1}} \frac{1}{\sqrt{2\omega'_1}}
 \frac{1}{\sqrt{2\omega_2}} \frac{1}{\sqrt{2\omega'_2}}\\
&\times\int \frac{d^3q}{(2\pi)^3} F_R(q) \frac{1}{{q^0}^2-\vec{q}\,^2-m_3^2+i\,\epsilon} t_{1} t_{2},\label{s2}
\end{split}
\ee
where $F_R(q)$ is the cluster form factor that we shall discuss below. Similarly, the full $S$ matrix for scattering of particle 3 with the cluster will be given by
\be
S=-i\, T \, (2\pi)^4 \delta(k+k_R-k'-k'_R)\times\frac{1}{{\cal V}^2}
\frac{1}{\sqrt{2 \omega_3}} \frac{1}{\sqrt{2 \omega'_3}}
\frac{1}{\sqrt{2\omega_R}} \frac{1}{\sqrt{2\omega'_R}}.\label{sful}
\ee

In view of the different normalization of these terms by comparing Eqs. \eqref{s11}, \eqref{s12}, \eqref{s2} and \eqref{sful}, we can introduce suitable factors in the elementary amplitudes,
\be
\tilde{t_1}=\frac{2M_R}{2m_1}~ t_1,~~~~\tilde{t_2}=\frac{2M_R}{2m_2}~ t_2,
\ee
with $m_1,~m_2,~M_R$ the masses of the particles 1,2 and the cluster, respectively,
where we have taken the approximations, suitable for bound states, $\frac{1}{\sqrt{2 \omega_i}}=\frac{1}{\sqrt{2m_i}}$, and sum all the diagrams by means of
\be
T=T_1+T_2=\frac{\tilde{t_1}+\tilde{t_2}+2~\tilde{t_1}~\tilde{t_2}~G_0}{1-\tilde{t_1}~\tilde{t_2}~G_0^2}. \label{new}
\ee
When $\tilde{t_1}= \tilde{t_2}$, as is the case here for the $\eta K$ and $\eta \bar K$ interaction, it can be simplified as
\be
T=\frac{2 \, \tilde{t_1}}{1-\tilde{t_1} \, G_0}. \label{new2}
\ee
The function $G_0$ in Eqs. \eqref{new} and \eqref{new2} is given by
\be
G_0(s)= \frac{1}{2 M_R} \int \frac{d^3\vec{q}}{(2\pi)^3} F_R(q) \frac{1}{q^{02}-\vec{q}^{~2}-m_3^2 +i\,\epsilon },
\label{gprop}
\ee
where $F_R(q)$ is the form factor of the cluster of particles 1 and 2. We must use the form factor of the cluster consistently with the theory used to generate the cluster as a dynamically generated  resonance. This requires to extend to the wave functions the formalism of the chiral unitary approach developed for scattering amplitudes. This work has been done in Refs. \cite{gamerjuan,yamajuan,Aceti:2012dd} for $S$-wave bound states, $S$-wave resonant states and states with arbitrary angular momentum, respectively. Here we only need the expressions for $S$-wave bound states, and then the expression for the form factors is given in section 4 of Ref. \cite{yamajuan} by
\begin{align}
\begin{split}
F_R(q)&=\frac{1}{\mathcal{N}} \int_{|\vec{p}\,|<\Lambda', |\vec{p}-\vec{q}\,|<\Lambda'} d^3 \vec{p} \; \frac{1}{2 E_1(\vec{p}\,)} \frac{1}{2 E_2(\vec{p}\,)} \frac{1}{M_R-E_1(\vec{p}\,)-E_2(\vec{p}\,)} \\
&\quad\frac{1}{2 E_1(\vec{p}-\vec{q}\,)} \frac{1}{2 E_2(\vec{p}-\vec{q}\,)} \frac{1}{M_R-E_1(\vec{p}-\vec{q}\,)-E_2(\vec{p}-\vec{q}\,)}, \label{formfactor}
\end{split}\\
\mathcal{N}&=\int_{|\vec{p}\,|<\Lambda'} d^3 \vec{p} \; \Big( \frac{1}{2 E_1(\vec{p}\,)} \frac{1}{2 E_2(\vec{p}\,)} \frac{1}{M_R-E_1(\vec{p}\,)-E_2(\vec{p}\,)} \Big)^2, \label{formfactorN}
\end{align}
where $E_1$ and $E_2$ are the energies of the particles 1, 2 and $M_R$ the mass of the cluster. The parameter $\Lambda'$ is a cut off that regularizes the integrals of Eqs. \eqref{formfactor} and \eqref{formfactorN}. This cut off is the same that one needs in the regularization of the loop function of the two particle propagators in the study of the interaction of the two particles of the cluster \cite{yamajuan}. As done in Refs. \cite{Yamagata2010,Xie:2010ig}, we take the value of $\Lambda'$ the same as the cutoff $q_{max}$ used to generate the resonance in the two-body interaction, which is the parameter to produce the cluster of $f_0(980)$ or $a_0(980)$ in our present work and will be discussed in the next section. Thus we do not introduce any free parameters in the present procedure.

In addition, $q^0$, the energy carried by particle 3 in the rest frame of the three particle system, is given by
\be
q^0(s)=\frac{s+m_3^2-M_R^2}{2\sqrt{s}}.
\ee

   Note also that the arguments of the amplitudes $T_i(s)$ and $t_i(s_i)$ are different, where $s$ is the total invariant mass of the three-body system, and $s_i$ are the invariant masses in the two-body systems. The value of $s_i$ is given by \cite{Yamagata2010}
\be
s_i=m_3^2+m_i^2+\frac{(M_R^2+m_i^2-m_j^2)(s-m_3^2-M_R^2)}{2M_R^2}, (i,j=1,2,\;i\neq j)
\ee
where $m_l \; (l=1,2,3)$ are the masses of the corresponding particles in the three-body system and $M_R$ the mass of two body resonance or bound state (cluster).

\section{$K \bar K$ and $\eta K\,(\eta' \bar K)$ two-body interactions}

To evaluate the Faddeev equations under the FCA, we need to define the two-body cluster and then let the third particle collide with the cluster. Thus, the starting point of our work is to look for the cluster in the two-body interactions. Following the formalism of Ref. \cite{Oller:1997ti}, by taking into account the chiral dynamics and the unitary coupled channels approach \cite{Kaiser:1995eg,Oller:1997ti, Oset:1997it,Oller:1998hw,Oller:1998zr,Oller:2000fj,Jido:2003cb,Guo:2006fu,Guo:2005wp,GarciaRecio:2002td, Hyodo:2002pk}, we should reproduce the resonances $f_0(980)$ and $a_0(980)$ as the cluster of FCA. We briefly summarize the method of Ref. \cite{Oller:1997ti} here.

To calculate the scattering amplitudes with the coupled channels unitary approach, the Bethe-Salpeter equation in coupled channels, with the factorized on shell potentials \cite{Oller:1997ti,Oller:2000fj} is used:
\be
t = [1-VG]^{-1} V,
\label{bese}
\ee
where the kernel $V$ is a matrix of the interaction potentials between the channels, given by \cite{Oller:1997ti}
\ba
V^{I=0}_{11} (s) &=& -\frac{1}{2f_\pi^2}(2s - m_\pi^2),\quad
V^{I=0}_{12} (s) = -\frac{\sqrt{3}}{4f_\pi^2}s,\quad
V^{I=0}_{22} (s) = -\frac{3}{4f_\pi^2}s,\\
V^{I=1}_{11} (s) &=& -\frac{1}{3f_\pi^2} m_\pi^2,\quad
V^{I=1}_{12} (s) = \frac{\sqrt{6}}{36f_\pi^2}(9s -8m_K^2 - m_\pi^2-3m_\eta^2),\,\,
V^{I=1}_{22} (s) = -\frac{1}{4f_\pi^2}s,
\ea
with $f_\pi$ the pion decay constant. Note that in isospin $I=0$, there are two coupled channels, 1 is $\pi \pi$ and 2 is $K \bar K$; for $I=1$, channel 1 denotes $\pi^0 \eta$ and 2 as $K \bar K$.

In Eq. (\ref{bese}) $G$ is a diagonal matrix of the loop function of two mesons in the {\it i}-channel, given by
\begin{equation}
G_i (s) = i \int\frac{d^{4}q}{(2\pi)^{4}}\frac{1}{(P-q)^{2}-m^2_1+i\varepsilon}\,\frac{1}{q^{2}-m^2_2+i\varepsilon},
\label{eq:G}
\end{equation}
where $m_1, ~m_2$ are the masses of the mesons in the $i$-channel, $q$ is the four-momentum of one meson, and $P$ is the total four-momentum of the system, thus, $s=P^2$. Note that the integral of Eq. \eqref{eq:G} is logarithmically divergent. Then, using a cut-off momentum to regularize it, we have
\begin{equation}
G_i (s) = \int_0^{q_{max}} \frac{d^3 \vec{q}}{(2\pi)^{3}}\frac{\omega_1+\omega_2}{2\omega_1\omega_2}\,\frac{1}{P^{0\,2}-(\omega_1+\omega_2)^2+i\varepsilon},
\label{eq:Gco}
\end{equation}
where $\omega_i = \sqrt{\vec{q}\,^2+m_i^2},~(i =1,\,2)$, and $q_{max}$ is the cut-off of the three-momentum, the free parameter. Also the analytic formula of Eq. \eqref{eq:Gco} can be seen in Refs. \cite{Guo:2005wp,Oller:1998hw}. On the other hand, the analytic expression of the dimensional regularization for Eq. \eqref{eq:G} can be seen in Ref. \cite{Oller:2000fj} (more discussions about $G_i$ are also seen in Refs. \cite{GarciaRecio:2010ki,Wu:2010rv,Xiao:2013jla}) with a scale $\mu$ fixed a priori and the subtraction constant $a(\mu)$ as free parameter.

\begin{figure}
\centering
\includegraphics[scale=0.6]{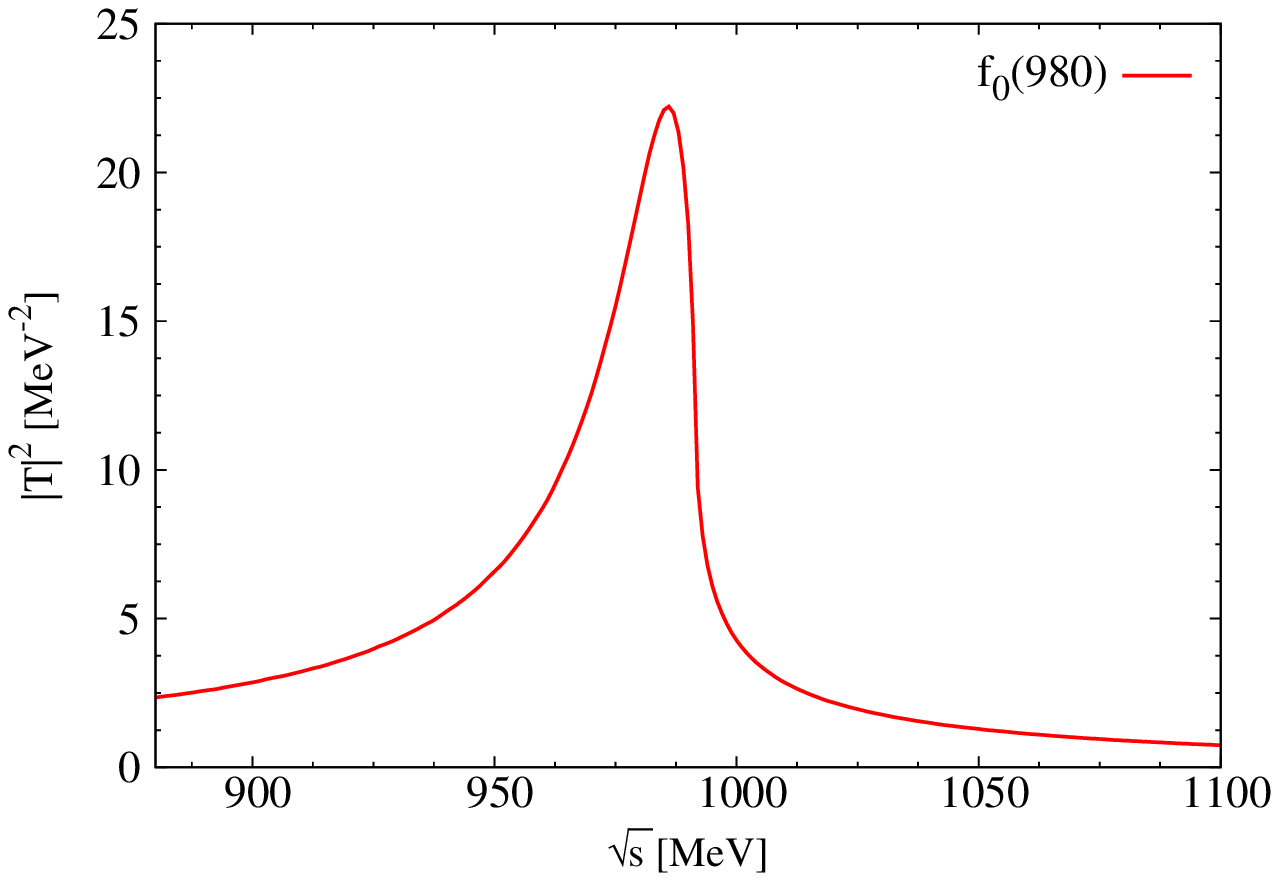}
\includegraphics[scale=0.6]{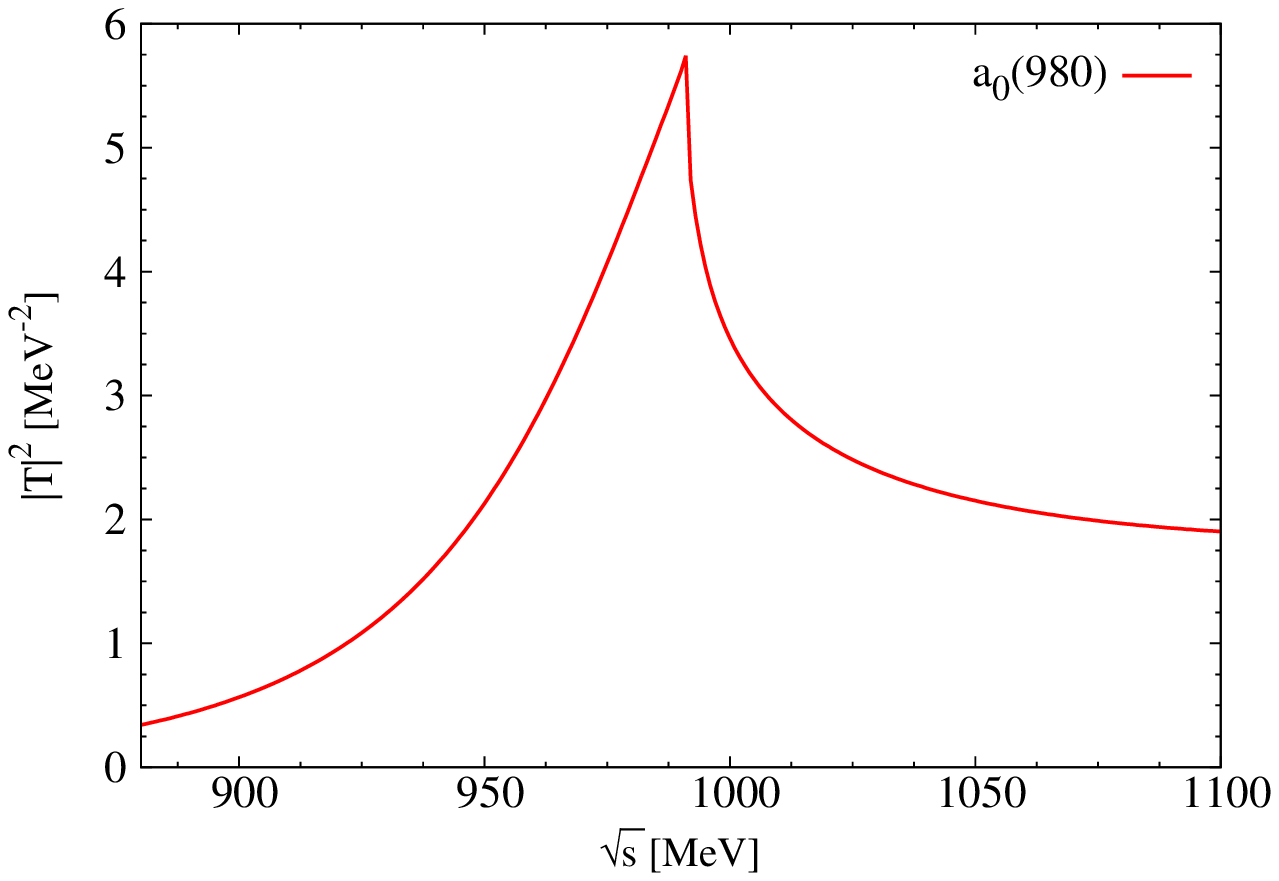}
\caption{Modulus squared of the scattering amplitudes. Left: $|t_{12}^{I=0}|^2/10^3,\; f_0(980)$; Right: $|t_{22}^{I=1}|^2/10^3,\; a_0(980)$. }\label{fig:tf0a0}
\end{figure}

Taking $\Lambda = 1.03 \gev$ and $\Lambda = \sqrt{q_{max}^2+m_K^2}$ as done in Ref. \cite{Oller:1997ti}, we get $q_{max}=903\mev$. Our results are shown in Fig. \ref{fig:tf0a0}.
In Fig. \ref{fig:tf0a0}, we produce the resonances of $f_0(980)$ and $a_0(980)$, which are consistent with Ref. \cite{Oller:1997ti} and form the clusters of the $\eta K \bar K$ and $\eta' K \bar K$ three-body interactions in our present work.

As discussed in the former section, to perform the evaluation of Faddeev equations under the FCA, we need the calculation of the two-body interaction amplitudes ($\tilde{t}_1$ and $\tilde{t}_2$) of $\eta K$ and $\eta \bar K$ for the $\eta K \bar K$ system ($\eta' K$ and $\eta' \bar K$ for the $\eta' K \bar K$ system) which are investigated in Refs. \cite{Oller:1998zr,Li:2002we,Guo:2005wp} as mentioned before. The former input is needed to construct the form factor of the cluster entering Eq. (\ref{gprop}).

\begin{figure}
\centering
\includegraphics[scale=0.6]{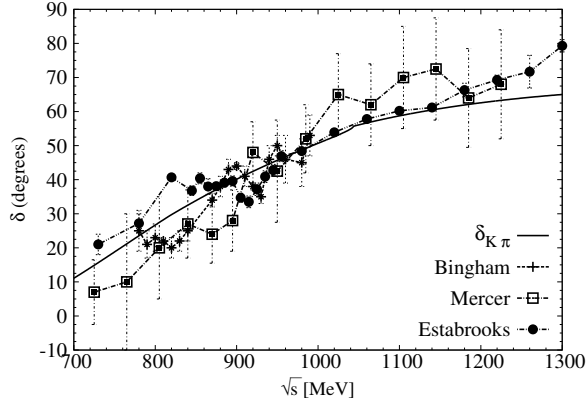}
\caption{The $S$-wave $K \pi$ phase shifts in isospin $I=1/2$. The experiments data are taken from:  Mercer \cite{Mercer:1971kn}, Bingham \cite{Bingham:1972vy}, Estabrooks \cite{Estabrooks:1977xe}.}\label{fig:phashi}
\end{figure}

Next we address the $\eta K$, $\eta \bar K$ and $\eta ' K$, $\eta ' \bar K$ interaction. Since we are involving the $\eta '$, it is convenient to take the three coupled channels $\pi K$, $\eta K$ and $\eta' K$, labeled by channel 1, 2 and 3 respectively. Thus, the potentials are \cite{Guo:2005wp}
\ba
V^{I=1/2}_{11} (s) &=& -\frac{1}{4f_\pi^2} (4s+3t-4m_\pi^2-4m_K^2),\\
V^{I=1/2}_{12} (s) &=& -\frac{\sqrt{2}}{6f_\pi^2} (-3t+2m_K^2+m_\eta^2),\\
V^{I=1/2}_{13} (s) &=& \frac{1}{12f_\pi^2} (-3t+3m_\pi^2+8m_K^2+m_{\eta'}^2),\\
V^{I=1/2}_{22} (s) &=& -\frac{2}{9f_\pi^2} (3t-m_K^2-2m_\eta^2),\\
V^{I=1/2}_{23} (s) &=& \frac{\sqrt{2}}{18f_\pi^2}(3t-3m_\pi^2 +2m_K^2 - m_\eta^2-m_{\eta'}^2),\\
V^{I=1/2}_{33} (s) &=& -\frac{1}{36f_\pi^2} (3t-6m_\pi^2+32m_K^2-2m_{\eta'}^2),
\label{v33}
\ea
where there is a minus sign difference with Refs. \cite{Oller:1998zr,migueleta} in some nondiagonal matrix elements resulting from taking different phase conventions \footnote{The final scattering amplitudes are the same, as pointed out by J. A. Oller.}. As done in Ref. \cite{Guo:2005wp}, we take
\be
 \mu = m_K,\qquad a(\mu)=-1.383, \label{eq:para1}
\ee
in the loop function for all channels, and we obtain the same results as in Ref. \cite{Guo:2005wp}, seen in Fig. \ref{fig:phashi}, which agree fairly well with the data except at the higher energies.

 With these parameters, we also find the pole of $\kappa$ [or $K^*(800)$], $(743.72-i275.36) \mev$, which is consistent with the result of Ref. \cite{Guo:2005wp}, $(0.742-i0.273) \gev$. Then, using these parameters, we can get the $\eta K$ and $\eta' K$ scattering amplitudes. Because of charge conjugation symmetry, the amplitudes for $\eta \bar K$, $\eta' \bar K$ are the same as those for $\eta K$, $\eta' K$.

\section{$\eta K \bar K$ and $\eta' K \bar K$ three-body interactions}

As discussed in the former section, we calculate the $\eta  K$ and $\eta \bar K$ ($\eta'  K$ and $\eta' \bar K$) amplitudes using the same parameters, and then we use Eq. \eqref{new2} to evaluate the three-body amplitude of the $\eta K \bar K$ ($\eta' K \bar K$) system. Also, as discussed  in the former section, the $\Lambda'$ of Eq. \eqref{formfactor} can be taken as $q_{max}=903\mev$ for the cluster of $f_0(980)$ or $a_0(980)$.

\begin{figure}
\centering
\includegraphics[scale=0.6]{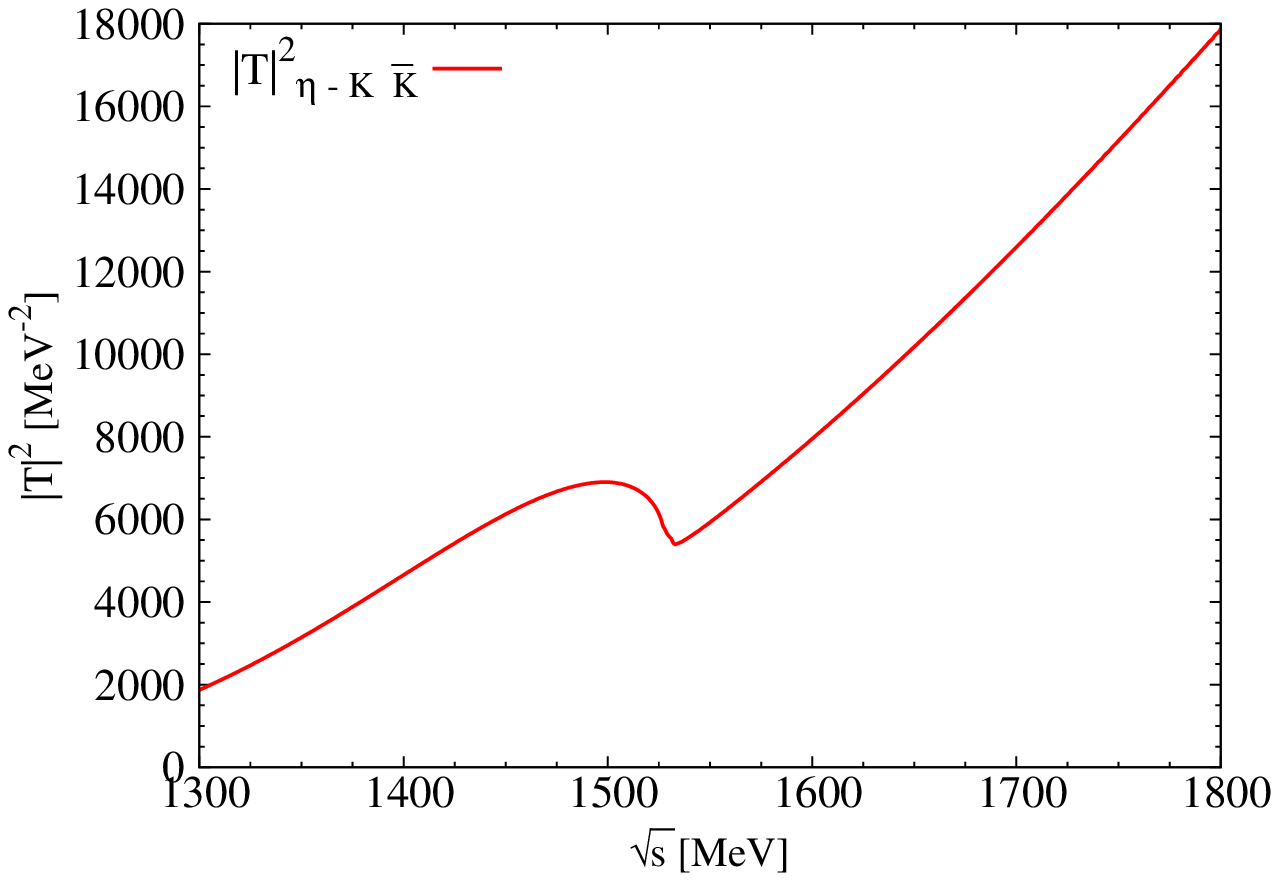}
\includegraphics[scale=0.6]{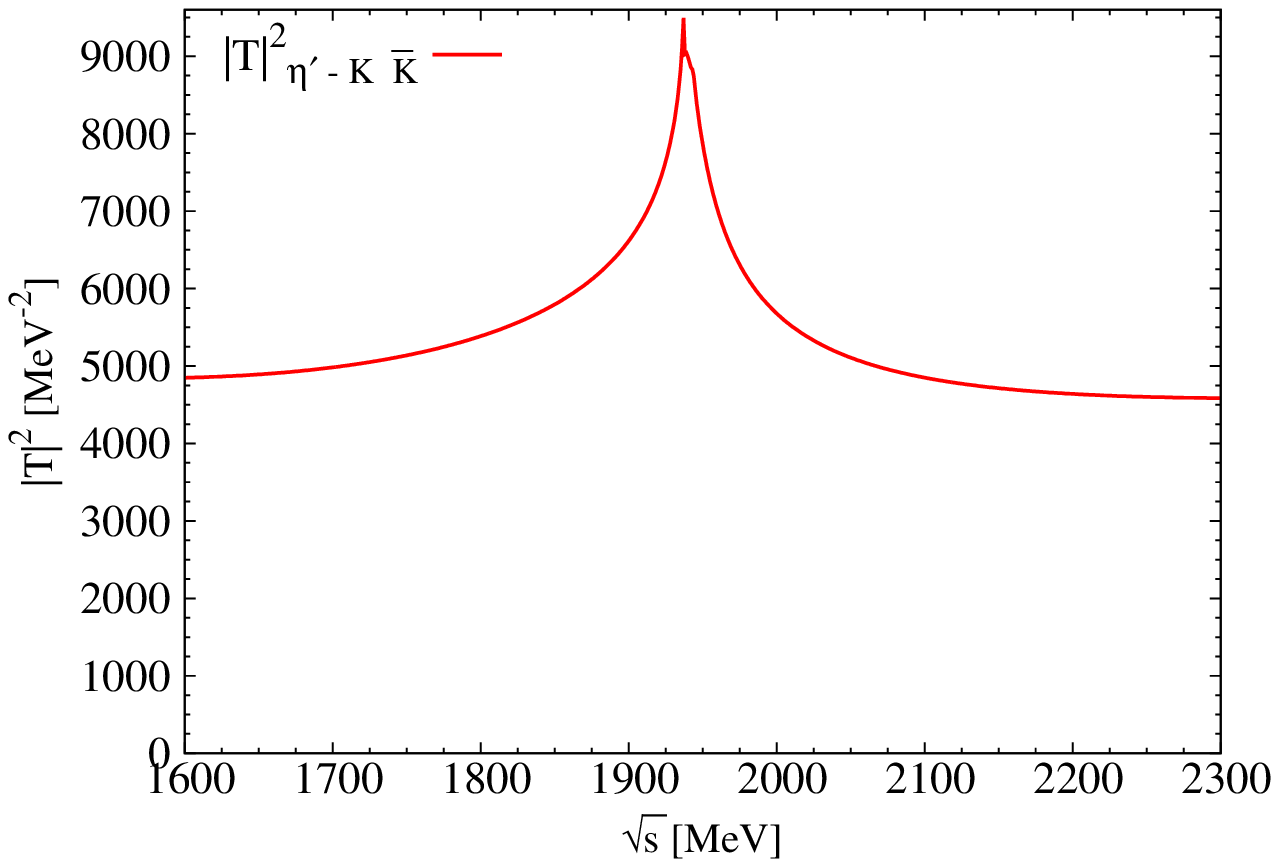}
\caption{Modulus squared of the three-body interaction amplitudes. Left: $|T_{\eta K \bar K}|^2$; Right: $|T_{\eta' K \bar K}|^2$.}\label{fig:tsqthree}
\end{figure}
In Fig. \ref{fig:tsqthree} (left), we can see a clear resonance structure in the modulus squared of the $\eta K \bar K$ scattering amplitude, which is around $1490\mev$, with the width of about $100\mev$, and about $38\mev$ below the threshold of $\eta\,f_0(980)$. This result is consistent with the one found in Ref. \cite{migueleta}. From the PDG \cite{pdg2012}, this resonance may be the $\eta(1475)$ of $I=0$, with mass $1476 \pm 4 \mev$ and width $85 \pm 9 \mev$. Comparing our results with the PDG, both the mass and the width are consistent with the experimental values if we assume 10-15 MeV uncertainties in our calculated results.

Since the masses of the $K \bar K$ bound states $f_0(980)$ and $a_0(980)$ are the same and only their isospins are different, the three-body amplitudes of $\eta K \bar K$ and $\eta' K \bar K$ in our formalism are degenerated in isospin $I=0$ and $I=1$. This means that if we predict a bound state for the $\eta f_0(980)$ system, we also have the same for $\eta a_0(980)$. This is so, assuming that the $f_0(980)$ and $a_0(980)$ resonances are predominantly $K \bar K$ molecules. But, as we have discussed, in the construction of the $f_0(980)$ resonance we need the $\pi \pi$ and $K \bar K$ channels, and the $\pi \pi$ is marginal in the structure of the resonance, it simply provides a decay mode. However, this is not the case for the $a_0(980)$ where the $\pi \eta$ channel already plays an important role in the build up of the resonance. Then a more elaborate, and technically complex, study of the $\eta, ~\eta '$ interacting with this system, would have much contribution from $\eta \eta$, which only comes from coupled channels and is very weak, and $\eta \pi$ which is also weak. The signal that we get in Fig. \ref{fig:tsqthree} would be much diluted and we do not expect an I=1 state.

We also see an obvious peak in Fig. \ref{fig:tsqthree} (right) for the $\eta' K \bar K$ interaction. But the mass position of the peak is about $1940\mev$, which is very close to  threshold, $1942\mev$. Therefore, this peak should be an enhancement effect of the threshold, a cusp effect, and we will check it further in the next section.

\section{Further discussions}

We showed the results of our investigation of the $\eta K \bar K$ and $\eta' K \bar K$ systems in the former section. For the $\eta K \bar K$ scattering, we find one resonance structure in the modulus squared of amplitude. But, for the other one, the clear peak of the $\eta K \bar K$ amplitude turns into an enhancement effect at the threshold in the $\eta' K \bar K$ amplitude, a cusp effect reflecting the cusp of the $\tilde t_1$ amplitude, used in Eq. (\ref{new2}), at threshold. In all these results we did not take into account the width of $f_0(980)$ as done in Ref. \cite{Xiao:2012dw}. In the PDG, the width of the $f_0(980)$ is 40 to 100 $\mev$, which is not small compared to the binding energy found.

\begin{figure}
\centering
\includegraphics[scale=0.6]{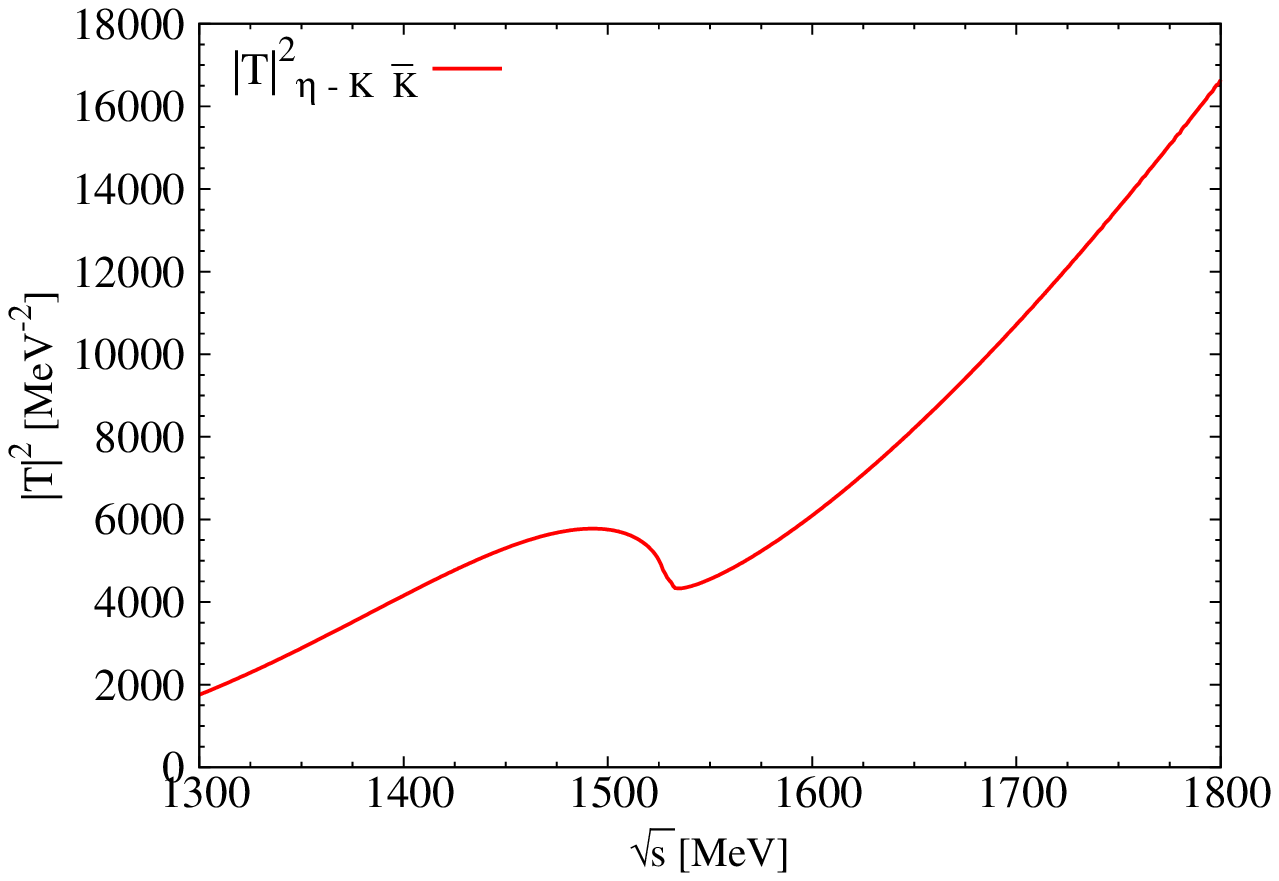}
\includegraphics[scale=0.6]{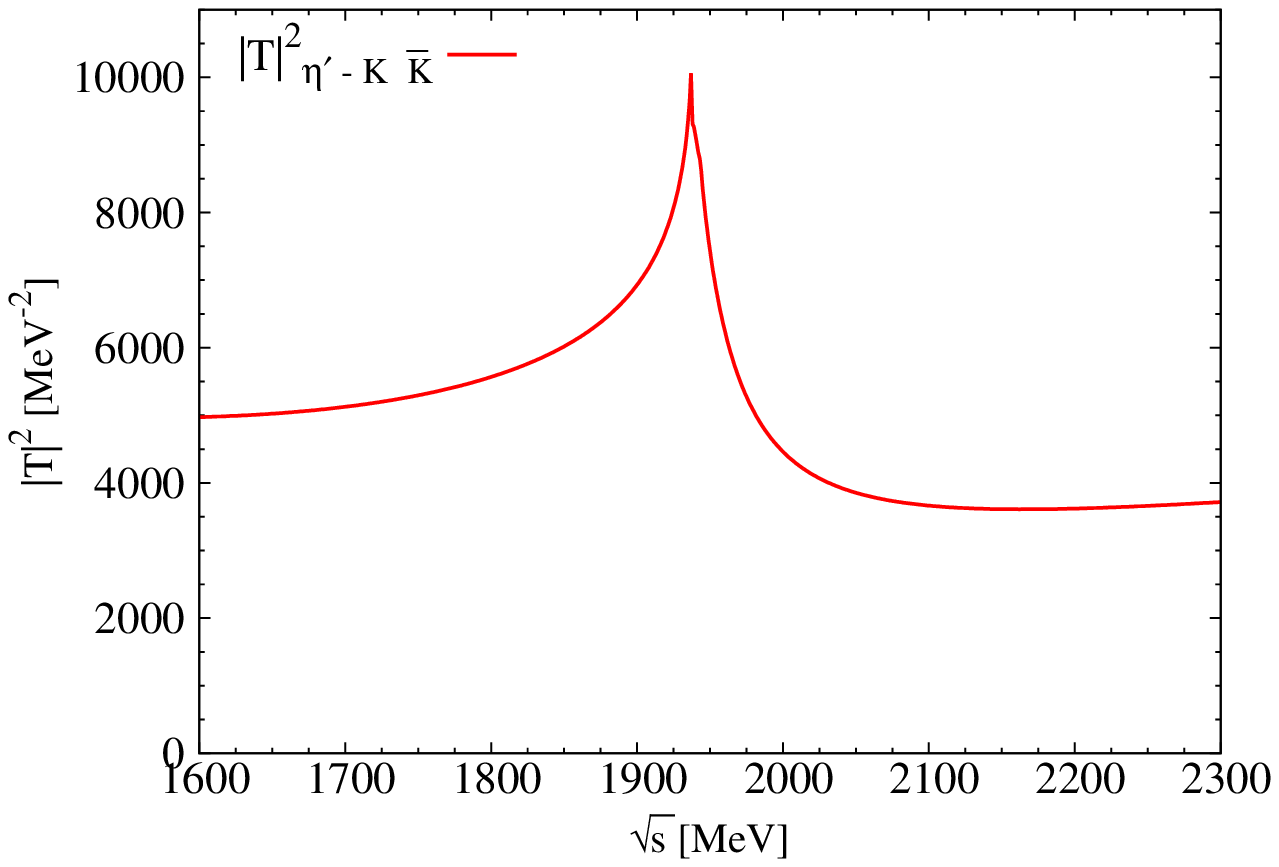}
\caption{Modulus squared of the three-body interaction amplitudes with the considering the contribution of the width of $f_0(980)$ [or $a_0(980)$]. Left: $|T_{\eta K \bar K}|^2$; Right: $|T_{\eta' K \bar K}|^2$.}\label{fig:tsqthree2}
\end{figure}

Following Ref. \cite{Xiao:2012dw}, we can take into account the width of the $f_0(980)$ in the three-body scattering amplitudes, just by replacing $M_R$ in Eqs. \eqref{formfactor}, \eqref{formfactorN} by $M_R-i\frac{\Gamma_R}{2}$. The new results are given in Fig. \ref{fig:tsqthree2}, where we just take the width as $60\mev$.
For the $\eta K \bar K$ amplitude we can see in Fig. \ref{fig:tsqthree2} (left), comparing with Fig. \ref{fig:tsqthree} (left), that the strength of the amplitude is reduced and the peak position is still not changed, but the width becomes a little larger (around 120 MeV), which is in the line with the finding in Ref. \cite{Xiao:2012dw}. For the $\eta' K \bar K$ amplitude, shown in Fig. \ref{fig:tsqthree2} (right), by comparing to  Fig. \ref{fig:tsqthree} (right), we can see that the strength at the peak is a bit increased and the shape changes a bit when considering the contribution of the width $f_0(980)$. The important thing, however, is that the shape of the  $\eta' K \bar K$ amplitude continues to be that of a cusp effect. In summary, as discussed in Ref. \cite{Xiao:2012dw}, we can conclude that the effects of the contribution of the cluster's width are small and do not change the relevant features found before.

\begin{figure}
\centering
\includegraphics[scale=0.6]{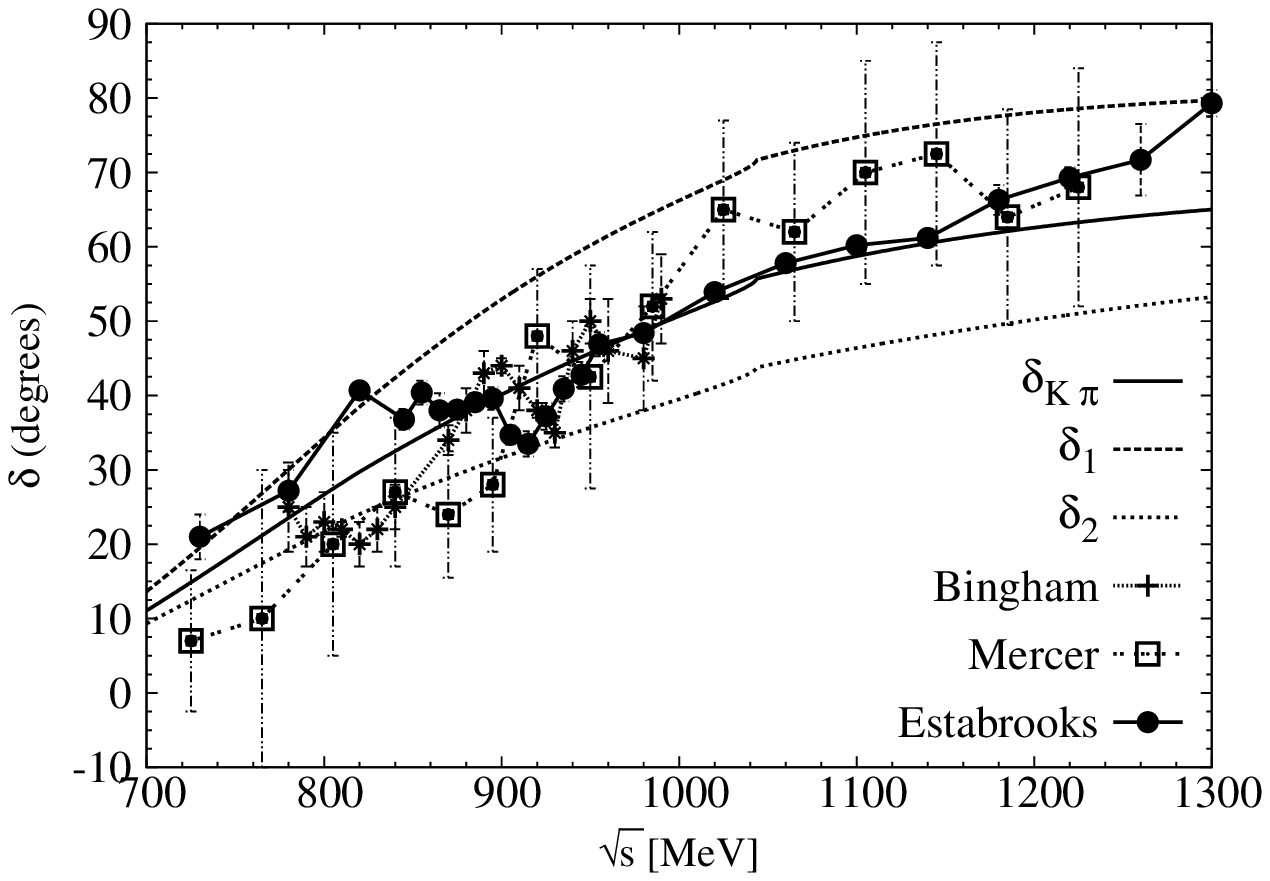}
\includegraphics[scale=0.6]{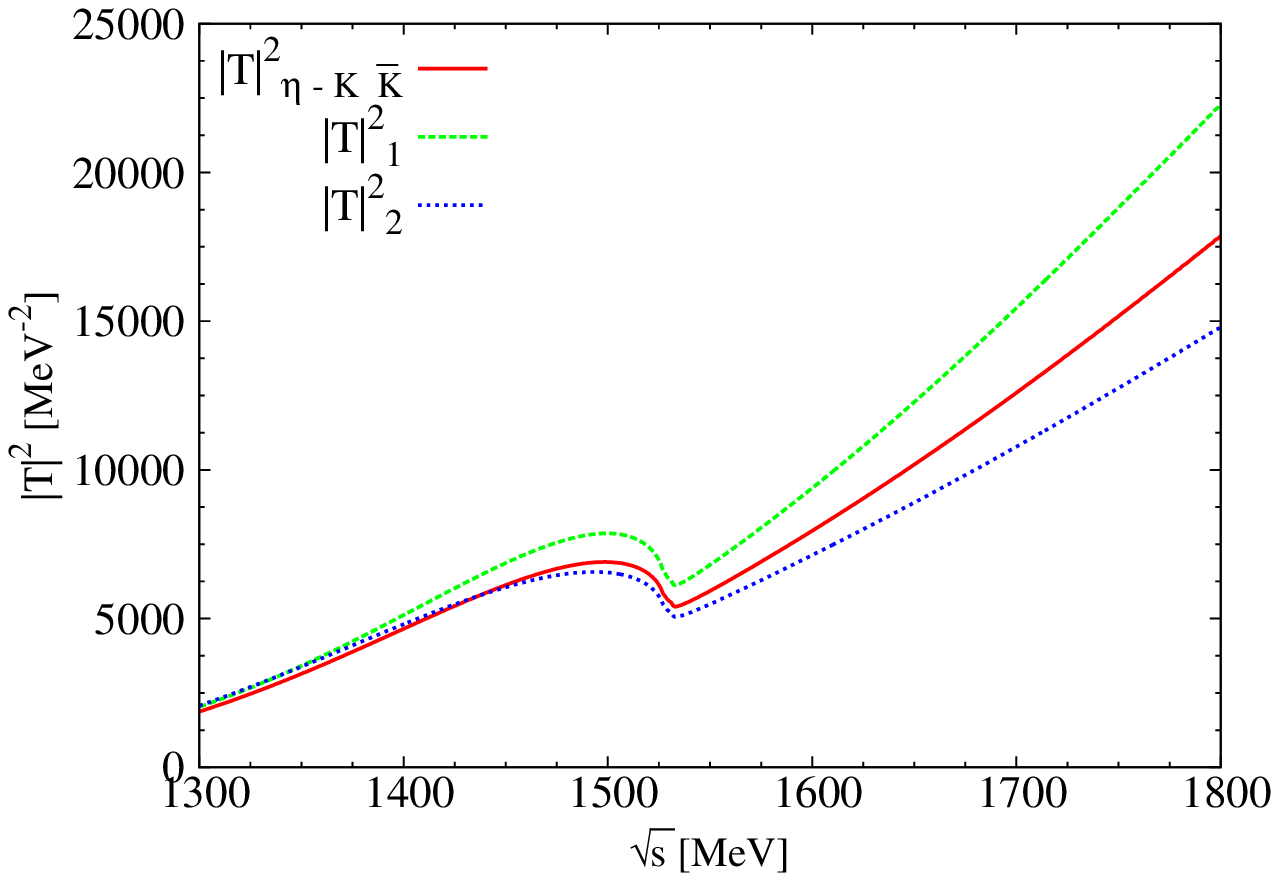}
\caption{The results for the change of the parameter in Eq. \eqref{eq:para1}. Left: the $K \pi$ phase shifts, solid line is the fit one, the two dash line of $\delta_1,~\delta_2$ are with 50 \% changes; Right: $|T_{\eta K \bar K}|^2$, solid line is with fit parameter, the two dash line of $T_1,~T_2$ are with 50 \% changes of the fit.}\label{fig:tsqthree3}
\end{figure}

Next, we want to check the uncertainties in Eq. \eqref{new2} when we make a small change in the parameters in the evaluation of $\tilde{t}_1$. Following Ref. \cite{Guo:2005wp}, we can only change $a(\mu)$. This parameter was chosen in Ref. \cite{Guo:2005wp} to fit the experimental data of the $K \pi$ phase shifts. Then, we change 50 \% up and down the parameter $a(\mu)$ of Eq. \eqref{eq:para1}, to a point where the $K \pi$ phase shifts are not too good, as shown in Fig. \ref{fig:tsqthree3} (left).
From Fig. \ref{fig:tsqthree3} (right), we can see that the resonance structure in $\eta K \bar K$ scattering is not changed so much even with these extreme changes in the input, and both the peak position and the width have practically not changed. This gives us confidence that the results that we get are rather solid and do not change with small variations of the parameters. The same changes only affect in a minor way the $\eta' K \bar K$ amplitude and the cusp effect at threshold is the only relevant feature of the amplitude.

  At this point one must comment on the results of \cite{migueleta}. In that work the interaction of the $\eta$ with the $K \bar K $ cluster is done as here, although the formalism seems rather different. There the primary amplitude for the $\eta$ interaction with the components of the cluster is evaluated and then the $\eta$ and the cluster propagate similarly to the propagation of the meson meson components in the $G$ function. The caveat is that while the regularization parameters are fitted to data on meson meson scattering, here one does not have this information for the scattering of $\eta$ and $f_0(980)$ and one must make assumptions on how this new loop is regularized. As a consequence, there is an element of uncertainty and usually what one makes is to assume that the interaction gives rise to a certain resonance to fix the parameters, although they are kept within a natural range; there is, hence, not a genuine prediction. In that work, the  $\eta f_0(980)$ gives rise to the $\eta(1475)$ as we have also claimed here. But the $\eta ' f_0(980)$ is claimed to produce the $X$(1835) resonance, something that our approach does not give.

The difference between the $\eta K \bar K$ and $\eta ' K \bar K$ systems could be qualitatively understood by recalling that the $\eta K$, together with the
$\pi K$ system, generate the broad $\kappa$ resonance, but the $\eta ' K$ amplitude has no structure around the $\eta ' K$ energies (up to the unavoidable cusp at threshold) and is small and smooth around these energies.  In order to see how far we are from creating a resonance structure in the $\eta ' K \bar K$ system, we artificially multiply $V_{33}$ of Eq. (\ref{v33}) by a factor and look at the $\eta ' K \bar K$ amplitude. We must multiply by a factor four the $V_{33}$ potential to see the peak move a bit ( by about 6 MeV ) below the threshold. Since the uncertainties of the model are by no means that large (we can accept about 20 \% uncertainties in the potentials), the former exercise tells us about the cusp character of the $\eta ' K \bar K$ amplitude is quite a stable result and we cannot associate a physical $\eta$ state to it.

We should comment on the paper \cite{albereta}, where using the Faddeev approach in the version of  Ref. \cite{alberone}, one peak in $|T|^2$ for $\pi K \bar K$ is found around 1400 MeV, which is associated to the $\pi(1300)$. It is mentioned there that the $\eta K \bar K$ system is also investigated and no clear signal is seen. The coupled channels approach used there contains $\pi K$ and $\eta K$ but not $\eta ' K$. We have checked that removing the $\eta ' K$ channel does not change qualitatively the $\eta K \bar K$ amplitude, although the distribution of $|T|^2$ in energy has a broader shape. Consideration of the $\eta ' K$ channel makes the energy distribution a little sharper. The fact that no clear peak for the $\eta K \bar K$ amplitude appears is somewhat unexpected, since one usually gets qualitative agreement between the FCA and the Faddeev calculations for bound states. For instance, the three body $\bar K N K$ scattering amplitude was calculated using the FCA to the Faddeev equations in Ref. \cite{Xie:2011uw} and the results of that work are in good agreement with the other theoretical works \cite{Jido:2008kp,MartinezTorres:2008kh} evaluated using variational and Faddeev approaches, respectively. The same can be said when one studies the $\bar K NN$ system in the FCA \cite{Bayar:2012rk} or in Faddeev calculations \cite{Ikeda:2010tk}, or variational calculations \cite{Dote:2008hw}. The $DNN$ system is another case of agreement between the FCA and variational calculation \cite{Bayar:2012dd}. We state the present situation and call for further calculations of the $\eta K \bar K$ system using different approaches in order to clarify the situation.

\section{Conclusions}

In our work, we study the three-body systems of $\eta K \bar K$ and $\eta' K \bar K$, by using the fixed center approximation to the Faddeev equations. The clusters of $f_0(980)$  for the fixed center approximation is successfully reproduced by the chiral unitary approach. With this approach, the experimental $S$-wave $K \pi$ phase shifts of isospin $I=1/2$ are also  well fitted. For the three-body scattering we find a resonant structure in the $\eta K \bar K$ scattering amplitude, which may correspond to the $\eta(1475)$ state for $I=0$. This finding is consistent with the result of Ref. \cite{migueleta}. We also make an estimation of our theory uncertainties for this state by taking into account the contribution of the cluster's width and reasonable changes in the free parameters, and we get stable results. As for the $\eta' K \bar K$ scattering, we only get an enhancement effect at the threshold in the modulus squared of the interaction amplitude and we can not claim that this can be associated to any resonance.

\section*{Acknowledgements}

We thank J. A. Oller, M. Albaladejo, F. K. Guo and A. Martinez Torres for useful discussions.
This work is partly supported by the Spanish Ministerio de Economia y Competitividad and European FEDER funds under the Contract No. FIS2011-28853-C02-01 and the Generalitat Valenciana in the program Prometeo, 2009/090. We acknowledge the support of the European Community-Research Infrastructure Integrating Activity Study of Strongly Interacting Matter(acronym Hadron Physics 3, Grant No. 283286) under the Seventh Framework Programme of the European Union.
This work is also partly supported by the National Natural Science Foundation of China under Grant No. 11165005 and by scientific research fund (201203YB017) of Education Department of Guangxi.

\end{document}